\documentclass[aps,pre,twocolumn]{revtex4}{
\usepackage{amsmath,amsfonts,bm,color,amssymb}
\usepackage{graphicx,epsfig,overpic,hyperref,ulem}

\newcommand{\dt}{{\Delta t}}

\newcommand{\refeq}[1]{Eq. [\ref{#1}]}
\newcommand{\col}[1]{{\color{black}{#1}}}

\newcommand{\kT}{ k_BT}

\newcommand{\visrat}{{\chi}}

\newcommand{\im}{{\mathrm i}}

\newcommand{\jj}{{\ell}}

\newcommand{\q}{{\nu}}

\begin{document}

\title{Curvature fluctuations of fluid vesicles reveal hydrodynamic dissipation within the bilayer}
\author{Hammad A. Faizi$^1$, Rony Granek$^2$ and Petia M. Vlahovska$^3$}

\affiliation{$^1$Department of Mechanical Engineering, Northwestern University, Evanston, IL 60208, USA\\
$^2$The Avram and Stella Goldstein-Goren Department of Biotechnology Engineering, and The Ilse Katz Institute for Meso and Nanoscale Science and Technology, Ben-Gurion University of The Negev, Beer Sheva 84105, Israel\\
$^3$Department  of Engineering Sciences and Applied Mathematics, Northwestern University, 60208, USA, email: petia.vlahovska@northwestern.edu
}


\begin{abstract}

The biological function of membranes is closely related to their  softness, 
 which  is often studied through the membranes' thermally-driven fluctuations.  The analysis commonly assumes thatthe relaxation rate of a pure bending deformation is determined by the competition between  membrane bending rigidity and viscous dissipation in the surrounding medium. 
Here, we reexamine this assumption and demonstrate that viscous flows within the membrane dominate the dynamics of bending fluctuations  of non-planar membranes with a radius of curvature smaller than the Saffman-Delbr\"uck length. Using flickering spectroscopy of giant vesicles made of DPPC:Cholesterol mixtures and pure diblock-copolymer membranes, we experimentally detect the signature of membrane dissipation in curvature fluctuations, and show that membrane viscosity can be reliably obtained from the short time behavior of the shape time correlations. The results indicate that the DPPC:Cholesterol membranes behave as a Newtonian fluid, while
polymer membranes exhibit more complex rheology. Our study provides physical insights into the time scales of curvature remodeling of biological and synthetic membranes.
\end{abstract}

\date{ \today}

\maketitle

\section{Introduction}
Bilayers assembled from lipids are the main structural component of the membranes that envelop and compartmentalize biological and synthetic cells \cite{Lipowsky:1991,Buddingh:2017,Rideau:2018}.
In living cells, membranes are dynamic structures that undergo continual morphological transformations involving dramatic changes in curvature e.g.,  budding, fission, and fusion \cite{Bassereau_2018,BONIFACINO2004153, mcmahon:2005,Chernomordik-Kozlov:2008, Kozlov:2023}. Lipid bilayers are easily bent by thermal and active forces and the resulting fluctuations are both of biological relevance, e.g., in membrane remodeling \cite{Lipowsky:2021}, adhesion \cite{Zidovska:2006,Fenz:2017}, nuclear shape dynamics \cite{Chu:2017,Jackson:2023}, and of
fundamental interest in soft matter physics  
\cite{Seifert:1997,Monzel:2016,Turlier:2019,Gupta:2021}. The canonical problem of thermally-driven curvature fluctuations of a membrane was considered in the pioneering work by Brochard and Lennon nearly 50 years ago \cite{brochard.1975}. In this now standard model, an undulation with wavenumber $q$ of an initially planar membrane, modeled as an incompressible interface with bending rigidity $\kappa$, is dissipated only by the viscosity of the surrounding fluid $\eta$ and 
relaxes exponentially with a rate $\kappa q^3/4\eta$. Notably, membrane viscosity does not affect the dynamics of the curvature fluctuations.

Departure from planar geometry dramatically changes the membrane dynamics  
\cite{Olla:2000,Rochal:2005,Henle:2010, woodhouse_goldstein_2012,Rahimi:2013, Rahimi_thesis,Sigurdsson:2016,Vlahovska:Stone,Sahu:2020},
  since 
 in-plane (shear) and out-of-plane (bending)  displacements are coupled \cite{Rochal:2005}. 
For a quasi-spherical vesicle, whose shape is described in terms of fluctuating spherical harmonic modes 
 $r_s(\phi,\theta,t)=R\left(1+f(\phi,\theta,t)\right), \,\, f=\sum f_{\jj m}(t) Y_{\jj m}(\phi,\theta)$, 
 the relaxation rate of a mode amplitude $f_{\jj m}$ is predicted to be \cite{Olla:2000,Rochal:2005,Vlahovska:Stone,Vlahovska:2019}
\begin{equation}
	\label{eqw}
	\omega(\jj, \kappa, \visrat_s)=\frac{\kappa}{\eta R^3}\frac{(\jj-1)\jj(\jj+1)(\jj+2)\left(\jj(\jj+1)+\bar\sigma\right)}{4 \jj^3+6 \jj^2-1+\left(4 \jj^2+4 \jj-8\right) \visrat_s}\,,
\end{equation}
where $\visrat_s=\eta_m/R\eta$ is a dimensionless membrane viscosity parameter, the ratio of the Saffman-Delbr\"uck length ($\eta_m/\eta$) to the vesicle radius $R$, $\bar\sigma=\sigma R^2/\kappa$ is the reduced membrane tension. 
Setting $\visrat_s=0$ reduces \refeq{eqw} to the result for a non-viscous area-incompressible interface \cite{Milner-Safran:1987} (an area-compressible membrane has been considered in \cite{Komura-Seki:1993}).
The Brochard-Lennon's result, $\omega(\jj)\simeq\frac{ \kappa }{4\eta R^3}\, \jj^3$, is only valid at short-wavelengths, $\jj\gg \chi_s$.
For $\chi_s\gg 1$, a new regime is predicted to emerge in the relaxation spectrum for long-wavelength undulations $1 \ll \jj \ll \chi_s$, in which the dissipation is dominated  by membrane viscosity, 
$\omega(\jj)\simeq \frac{ \kappa }{4\chi_s\eta R^3}\, \jj^4$ . 
  This suggests that membrane viscosity can be deduced from the relaxation rates of the curvature fluctuations at equilibrium.  This approach is guaranteed to be in the linear response regime, unlike some of existing methods which rely on externally imposed  perturbations \cite{Faizi:2022, Dimova:1999, Tam:2021}. 

The prerequisite for pure bending mode damped by flows in the membrane rather than the bulk fluid, $\chi_s\gg 1$, 
is met if either the membrane viscosity is large, $\eta_m\gtrsim\eta R$, as in diblock-copolymer bilayers \cite{Dimova2002}, or vesicle size is small, $R\lesssim \eta_m/\eta$, as in submicron lipid liposomes. 
In this work, we report experimental evidence of membrane viscous dissipation in the  flickering of giant vesicles. We theoretically analyze the shape fluctuations of a quasi-spherical vesicle and derive the experimental observables that are sensitive to dissipation: the transverse mean square displacement of a membrane segment and the time-averaged autocorrelation function of the Fourier modes representing the contour of the equatorial cross-section. 
We find that the  latter decays at short times as a stretched exponential, with a universal stretching exponent 3/4, when membrane viscosity dominates dissipation, and approaches single exponential relaxation at long times. Combining the static and dynamic fluctuations spectra allows to  measure independently the bending rigidity and the membrane viscosity.  We apply this approach to characterize the viscous dissipation in bilayers made of poly(butadiene)-$b$-poly(ethylene oxide) diblock copolymers or lipid bilayers in the liquid ordered state  (mixtures of dipalmitoylphosphatidylcholine (DPPC) and cholesterol (Chol)), for which very limited data exists.

\section{Results and discussion}

\begin{figure*}
	\includegraphics[width=\linewidth]{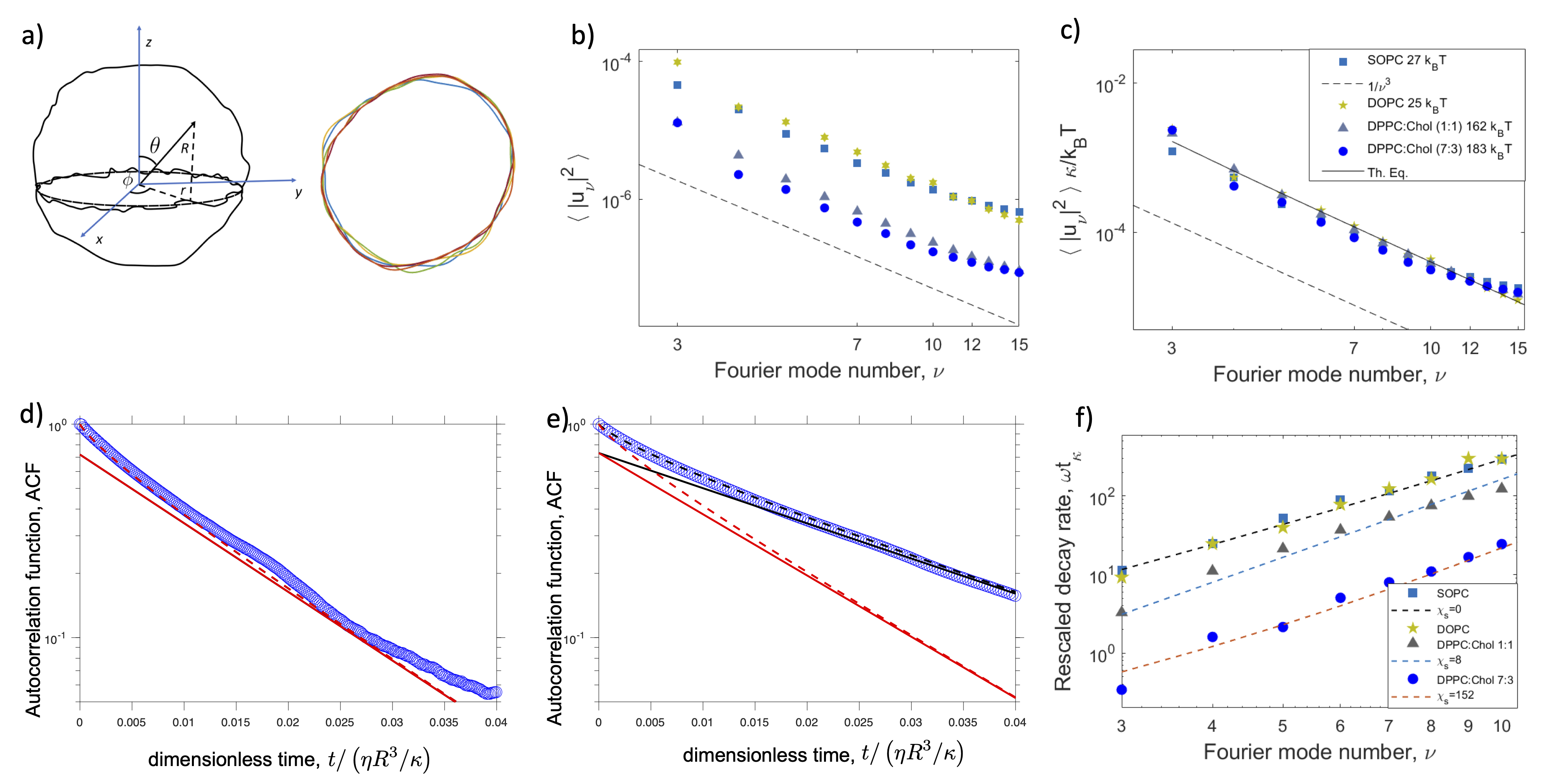}
	\caption{\footnotesize{(a) Sketch of a quasi-spherical vesicle and of time-lapse vesicle contours in the equatorial plane. (b) Power-spectrum of the contour fluctuations yields the bending rigidity $\kappa$. (c) The rescaled  static power spectrum by $\kappa$ is a universal function of the wavenumber. Solid line is \refeq{ACF} with $t=0$.
			(d) and (e): Autocorrelation functions (ACF) for Fourier mode 6 of the fluctuating equatorial contour of vesicles made of SOPC and  DPPC:Chol (1:1). Blue symbols are the experimental data. Dashed lines are the full theory \refeq{ACF} and the solid lines are the single exponential decay with rate given by \refeq{eqw}.   Red and black line colors correspond to dimensionless membrane viscosity $\visrat_s=0$ and $\visrat_s=8$, respectively.  			The time scale $t_\kappa=\eta R^3/\kappa$ is 23.3 s in (d) and  7.6 s in (e). 
(f) The long-time single exponential decay rate, rescaled by the bending relaxation time, obtained from the ACF as a function of the mode number. Dashed line is the theory \refeq{eqw}. }}
	\label{fig1}
\end{figure*}

\subsection{Autocorrelation function of the thermally excited membrane undulations}

In flickering experiments \cite{Gracia:2010,Faizi:2020}, a time series of the equatorial cross-section of a giant quasi-spherical vesicle (radius $R\sim 10 \,\mu$m)
is recorded. The quasi-circular 
contour 
 is decomposed in Fourier modes, $r_s(\phi,\pi/2,t)=R\sum u_{\q}e^{-\im \q\phi}$.   Their  autocorrelation function (ACF)  picks up all the $m=\q$ terms of the expansion of the vesicle shape  in spherical harmonics (see Appendix), leading to an ACF in the form
\begin{equation}
\begin{split}
	\mathrm{ACF}&(t)=\langle  u_{\q} (0)  u_{\q}^*(t) \rangle=\\
	&\sum_{\ell=|\q|}^{\ell_{\max}}  \frac{\kT}{\left((\jj-1)(\jj+2)\left(\jj(\jj+1)\kappa+\bar \sigma\right)\right)}
	 n^2_{\ell\nu}\Big|P_{\ell \q}(0)\Big|^2 e^{-\omega(\jj)t}
	 \end{split}
	\label{ACF}
\end{equation}
where $\kT$ is the thermal energy (k$_B$ is the Boltzmann constant and T is the temperature),  $P_{\ell \q}$ is the associate Legendre polynomial and $n_{\jj \q}$ is a normalization factor (see for definitions Appendix and \refeq{Ylm}). 

The mean square amplitude (static spectrum) of the fluctuations, $\langle |u_{\q}|^2 \rangle$, obtained from \refeq{ACF} by setting $t=0$, depends only on the membrane elastic properties (bending rigidity and tension). Indeed, the fluctuations spectrum shown in Figure \ref{fig1}b follows bending-dominated scaling \cite{Pecreaux:2004,Gracia:2010} (see also Appendix), $\sim 1/\nu^3$. 
Rescaling the spectrum by the bending rigidity collapses the data, see Figure \ref{fig1}c, and confirms that the static spectrum is controlled solely by bending rigidity. 

The decay of the ACF depends  on the membrane viscosity and thus can serve as a reporter for dissipation due to in-plane shear flows in the membrane if $\visrat_s$ is large enough.  DOPC and SOPC viscosities are reported to be  4.1$\pm2.6$ nPa.s.m and 9.7$\pm$5.8 nPa.s.m, respectively \cite{Faizi:2022}, corresponding to dimensionless surface viscosities $\visrat_s\lesssim 1$  for a typical 10 $\mu$m GUV, too small to have a detectable effect on vesicle shape fluctuations, see Figure \ref{fig1}d.  To achieve a lipid bilayer with high viscosity, we choose (i) DPPC:Chol mixtures, because they are in the liquid-ordered phase and thus expected to be very viscous  \cite{WANG20162846,Faizi:2022}, and (ii) di-block copolymers composed of hydrophilic and hydrophobic blocks, which are known to be very viscous \cite{Rumy:2002}.
Membrane viscosity measured with the electrodeformation method \cite{Faizi:2022, SI}  yielded 57.6$\pm$12.6 nPa.s.m  for DPPC:Chol (1:1),  83.6$\pm$14.3 nPa.s.m for DPPC:Chol (6:4), 1450$\pm$928 nPa.s.m for DPPC:Chol (7:3), $14.4\pm 4.4$ nPa.s.m for PS0 and $686\pm51$ nPa.s.m for PS1, spanning a range of dimensionless viscosities, $\sim 1-150$. 
Figure \ref{fig1}e  demonstrates that indeed the curvature fluctuations of the DPPC:Chol mixtures relax much more slowly compared to SOPC, indicating significant membrane viscosity, in this case by a factor of 10 larger than the SOPC viscosity.

Obtaining the  bending rigidity and tension  from the static spectrum leaves only one unknown parameter, the membrane viscosity, in the ACFs of the Fourier modes. The long-time exponential decay can be easily fitted, see solid lines in Figure \ref{fig1}de.
 Using this value of the membrane viscosity in \refeq{ACF} describes well the full relaxation curve (dashed lines). 
 Figure \ref{fig1}f summarizes the relaxation rates obtained from the long time single exponential fit of mode numbers 3-10. The slowing down of the ACF decay suggests
 that DOPC:Chol mixtures are much more viscous than the bilayers in the liquid disordered phase, DOPC and SOPC. The dependence of the relaxation rates on mode number is well described by \refeq{eqw} and yields the membrane viscosity.

\begin{figure}[h]
\includegraphics[width=\columnwidth]{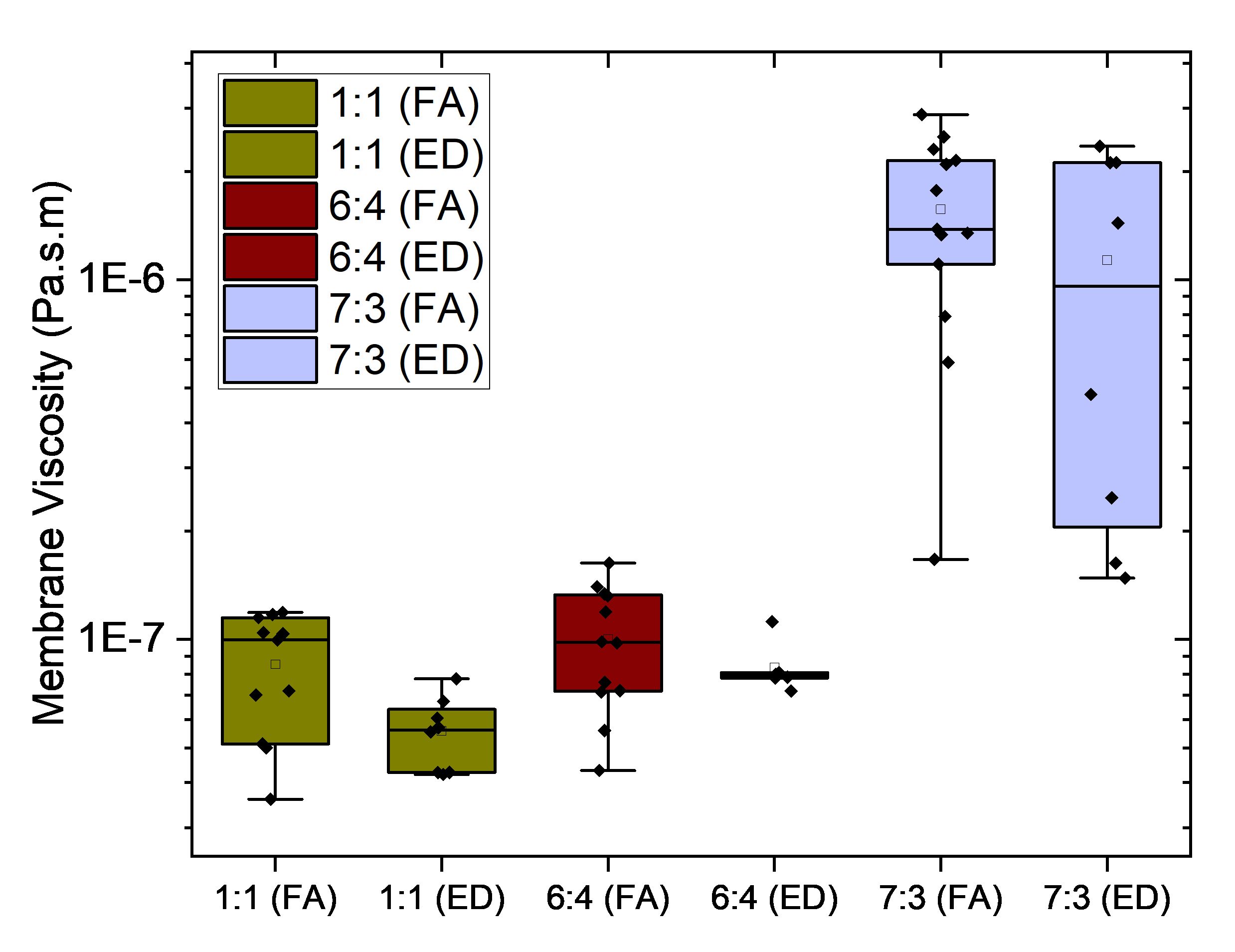}
\centering
\caption{\footnotesize{Membrane viscosity obtained from flickering spectroscopy (FA) and electrodeformation (ED) for different  DPPC: Chol mixed bilayers.}}
\centering
\label{figV}
\end{figure}

The viscosity obtained from the ACF fits agrees well with the data obtained from electrodeformation, see Figure \ref{figV}.  Membrane viscosity increases sharply with the DPPC fraction, demonstrating that the mixed membrane viscosity is not an additive property of the single-component bilayer properties, as also reported for other mixed systems \cite{Kelley:2020}.

\subsection{Subdiffusive fluctuation dynamics}

The ACF of the equatorial Fourier mode of order $\nu$, \refeq{ACF}, is not a single exponential, unless at sufficiently long times $\omega(\nu) t\gtrsim 1$ where $\text{ACF}\sim \text{e}^{-\omega(\nu) t}$. 
At short times, $\omega(\nu) t\ll 1$, all modes with $\jj \geq |\nu|$ contribute and the ACF decay is non-exponential,
effectively approaching a stretched exponential decay $\sim\exp\left[-(\gamma(\nu)t)^\zeta\right]$ with
{stretching exponent $\zeta$} and relaxation rate $\gamma(\nu)$.
More precisely (see Appendix for details), there are two ``short time" regimes (i.e. for $t\ll \omega(\nu)^{-1}$), commencing by a regime where dissipation is dominated by the solvent viscosity followed by a membrane viscosity dominated regime. The crossover time separating the two regimes is
\begin{equation}
\label{ctime}
	t^{*}\approx \frac{4\eta R^3}{\kappa \chi_s^3}\; .
\end{equation}

For early times, $t_0 \ll t \ll t^{*}$, where $t_0$ is the shortest relaxation time, $t_0=1/\omega(\jj_{\max})\sim \eta d^3/\kappa$, where $d$ is the bilayer thickness,  we get for the ACF in the solvent dissipation regime
\begin{equation}
	\langle u_\nu (0) u_\nu^*(t) \rangle \simeq   
	 \langle|u_\nu|^2\rangle - \frac{1}{ 8\pi^2} \frac{k_BT}{ \eta R^3}\; t \left[{1\over 3}\ln\left(\frac{4\eta R^3}{\kappa \nu^3\, t}\right) + 0.2838\right] 
	\label{ACF-asymp2}
\end{equation}

For later times, $t^{*}\ll t\ll \omega(\nu)^{-1}$ -- given that such a regime can be manifested, i.e. for $\chi_s\gg \nu$ --  we obtain that the ACF in the membrane dissipation regime is approximated by
\begin{equation}
	\langle u_\nu (0) u_\nu^*(t) \rangle\simeq   
		\langle|u_\nu|^2\rangle - {\Gamma[1/4]\over 6\pi^2} \frac{k_BT}{\kappa} \left(\frac{ \kappa } {4\chi_s\eta R^3}\; t\right)^{3/4} 
	\label{ACF-asymp1}
\end{equation}
presenting a non-exponential relaxation of the ACF from its static value; it may be cast as
\begin{equation}
	\Big< u_\nu (0) u_\nu^*(t) \Big>\simeq   
		 \big<|u_\nu|^2\big>\left(1-\left(\gamma(\nu) t\right)^\zeta\right)
	\label{ACF-asymp3}
\end{equation}
where $\zeta=3/4$ and the effective relaxation rate $\gamma(\nu)$ is
\begin{equation}
	\gamma(\nu)= {\Gamma[1/4]^{4/3}\over 4} \frac{ \kappa\, \nu^4} {\chi_s\eta R^3}= {\Gamma[1/4]^{4/3}\over 4} \frac{ \kappa\, \nu^4} {\eta_m R^2}\;,
	\label{ACF-gammaq}
\end{equation}
such that $\gamma(\nu)= \Gamma[1/4]^{4/3}\omega(\nu)\sim\nu^4$. 

\begin{figure*}
\centering
	\includegraphics[width=\linewidth]{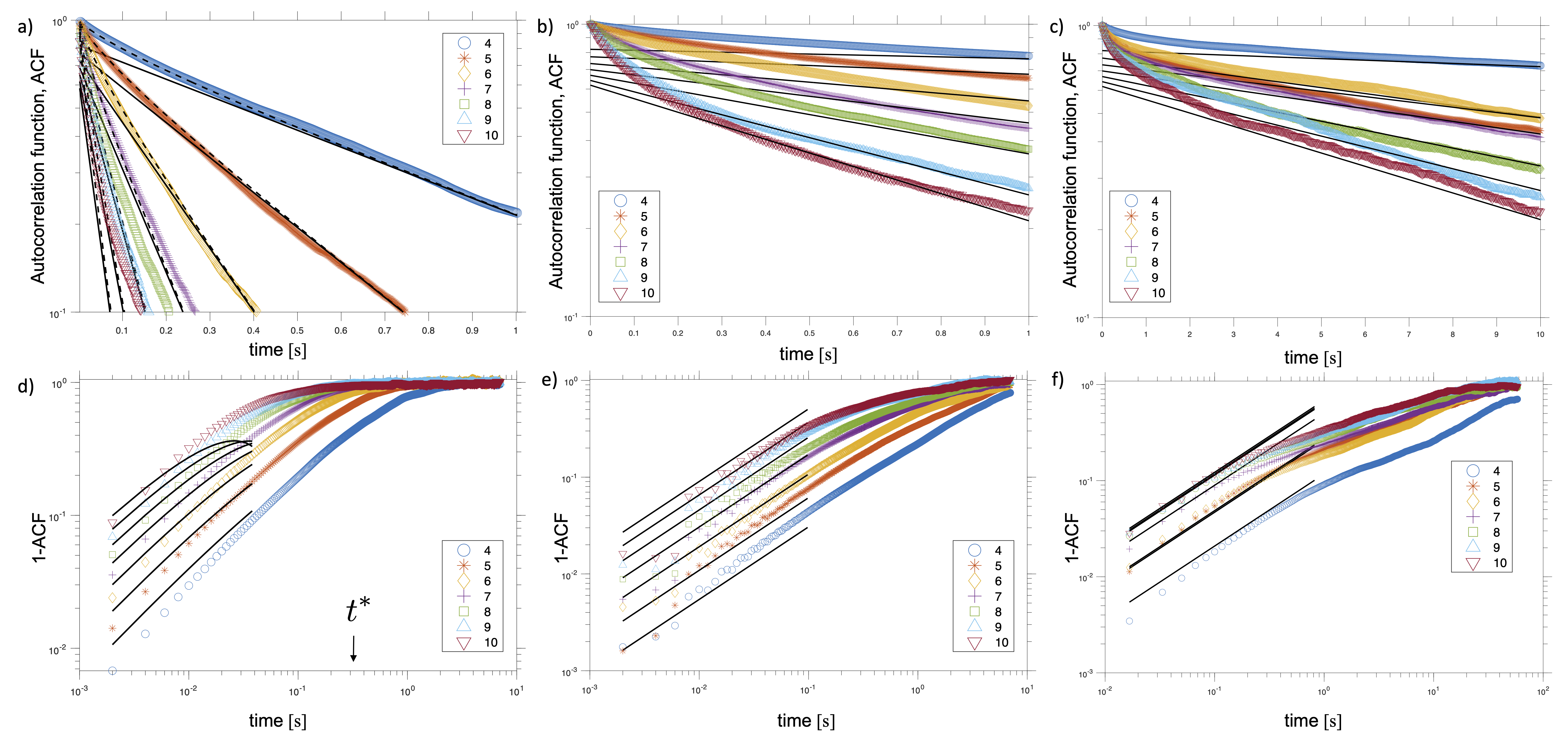}
	\caption{\footnotesize{
		\col{Normalized} autocorrelation functions (ACF) for Fourier modes 4-10 of the fluctuating equatorial contour of vesicles made of  DPPC:Chol (1:1) (a,d)  DPPC:Chol (7:3) (b,e), and PS1 (c,f). Symbols are the experimental data. In (a)-(c) the dashed lines are the full theory \refeq{ACF} and the solid lines are the single exponential decay with rate given by \refeq{eqw}. 
  (d)-(f) zoom into the short-time behavior of the ACF. The solid lines are the non-viscous-membrane asymptote \refeq{ACF-asymp2} in (d) and the viscous-membrane asymptote \refeq{ACF-asymp1} in (e) and (f).
  $t^*$ denotes the crossover time from relaxation dominated by bulk viscosity to membrane viscosity \refeq{ctime}.}}
\label{fig2}
\end{figure*}

The mean square displacement  of a Fourier mode $\nu$ is directly related to the ACF, $2(\langle|u_\nu|^2\rangle - \langle u_\nu (0) u_\nu^*(t) \rangle$ (see Appendix for the derivation). Thus the
``stretching" exponent, $\zeta=3/4$, is equivalent to the anomalous diffusion exponent of equatorial Fourier modes. It is identical to the one governing rod-like semi-flexible polymers obeying the worm-like chain model. This is interesting and can be rationalized by the following argument. First, by looking at the Fourier modes of deformations at the equator, the effective dimensionality of the Helfrich bending energy phase space is reduced from two to one, as for semi-flexible polymers  \cite{Farge:1993,Granek:1997}. Second, the membrane viscosity-dominated dissipation suppresses the long-range solvent hydrodynamic interaction, which again leads to a similar behavior as in rod-like semi-flexible polymers where the effect of the solvent-mediated hydrodynamic interaction is marginal (logarithmic).

To summarize, if membrane viscosity is dominating the relaxation and mode number is high (such that $\chi_s\gg \nu\gg 1$), we find that the ACF relaxation profile  can be approximated by the following form 
\begin{equation}
    \langle u_\nu (0) u_\nu^*(t) \rangle=   
		 \big<|u_\nu|^2\big> U_{\nu}[\col{\gamma(\nu)} t]
\label{ACF-scaling-assumption}
\end{equation}
where the scaling function $U_{\nu}(y)$ (for $\nu\gg 1$) obeys
\begin{equation}
    U_{\nu}(y)\simeq \begin{cases}
        1- y^{3/4}\;\;\;\;\;\;\;\;\;\;\;\;\;\;\;\;\;\;\;\;\;\;\;\;\;\;\;\;\;\;\; \text{for}\;\; y\ll 1\;,\cr
        \frac{\text{const.}}{\nu^{1/2}} \text{exp}\left[-y/\col{\Gamma[1/4]^{4/3}}\right] \;\;\;\;\;\text{for}\;\; y\gg 1\
    \end{cases}
\label{ACF-scaling-function}
\end{equation}
($\text{const.}\simeq 5.32$).

Figure \ref{fig2} compares the experimental ACFs for DPPC:Chol  and PS1 systems and the theoretical predictions. The DPPC:Chol (1:1) is only moderately viscous and the long-time single-exponential fit of the ACFs in Figure \ref{fig2}a yield dimensionless viscosity $\visrat_s=5$, corresponding to membrane viscosity $85$ nPa.s.m. Accordingly, the short time relaxation of the ACF (and the MSD), at times shorter than the crossover time $t^*$, is dominated by the dissipation in the solvent.  
The DPPC:Chol  (7:3) and the PS1 membranes are much more viscous. The long-time single exponential fits of the modes ACFs yield much higher dimensionless membrane viscosity,
$\visrat_s>100$.  Accordingly, their ACFs show clear $t^{3/4}$ power-law decay at short times. 
The fit of the short-time behavior with \refeq{ACF-asymp1} yields viscosity  $\visrat_s=150$ for the DPPC:Chol (7:3) membrane. However, the viscosity deduced from the long-time exponential decay of the ACF of each mode shows a trend to increase with the mode number, from $\visrat_s=180$ to 450. This more complex long-time dynamics may be a result from additional dissipation 
due to interpenetrating hydrophobic blocks of the two leaflets \cite{Seifert-Langer:1993,Watson:2010,Watson:2011}.
The intermonolayer friction effect becomes more pronounced at shorter wavelenghts thereby manifesting itself as a mode-dependent viscosity.  This effect is absent in the current model, \refeq{eqw}, which considers the membrane to be a structureless interface. 
Another possibility is diffusional softening in mixed bilayers \cite{SodtBJ:2022,Sapp-Sodt:2023,Hossein-Sodt:2024} originating from a  dynamic coupling between
the lateral distribution of lipids with differing curvature preference and the membrane
undulations.

The unique value for the viscosity obtained from the short-time fit with the ACF asymtptote suggests it  is the more reliable value. 

\subsection{Transverse subdiffusion of a membrane segment and dynamic structure factor of a vesicle membrane}
While flickering spectroscopy measures the meas square displacement of the Fourier modes,  scattering techniques such as neutron spin echo
\cite{Nagao:2017}, dynamic light scattering \cite{Freyssingeas:1997}, X-ray photon correlation spectroscopy \cite{Falus:2005} and some fluctuations experiments   \cite{Betz:2012,Helfer:2001a} measure
dynamic structure factor, $S(k,t)$, that is controlled by the single-point membrane mean square displacement (MSD), $\langle (\Delta h(t))^{2}\rangle$,
$S(k,t)\sim \text{Exp}[-\frac{k^2}{2}\langle (\Delta h(t))^{2}\rangle]$ \cite{Zilman-Granek:1996, Watson:2011, Zilman-Granek:2002,Granek:EPJE}. Hence, it is instructive to consider the effect of membrane viscosity on measurements made by these methods.

For planar membranes and non-viscous vesicles, the transverse (i.e. normal) membrane MSD, $\langle (\Delta h(t))^{2}\rangle\equiv R^2 \langle (\Delta f(t))^{2}\rangle$, follows the well-known prediction by Zilman and Granek (ZG)\cite{Zilman-Granek:1996,Zilman-Granek:2002},
\begin{equation}
		\langle (\Delta h(t))^{2}\rangle\simeq 
		\frac{\Gamma[1/3]}{2\pi } \frac{k_B T}{\kappa^{1/3}} \left(\frac{t}{4\eta}\right)^{2/3} \,.
	\label{ZG-asymptotes}
\end{equation} 
In contrast, for viscous vesicles, the ZG behavior is limited to earlier times, \col{$t_0\ll t\ll t^*$}, and is absent altogether if $\visrat_s\gtrsim R/d$. For longer times, \col{$t^*\ll t\ll \eta_m R^2/\kappa$} and $\jj_{max}\gg 1$ \cite{Granek:EPJE},
 the MSD asymptotically exhibits subdiffusive behavior with exponent $1/2$:
\begin{equation}
		\langle (\Delta h(t))^{2}\rangle\simeq 
\frac{Rk_BT}{4\sqrt{\pi}} \left(\frac{t}{\kappa \eta_m}\right)^{1/2} \,,
	\label{MSD-asymptotes}
\end{equation}

\begin{figure}[h]
\includegraphics[width=\columnwidth]{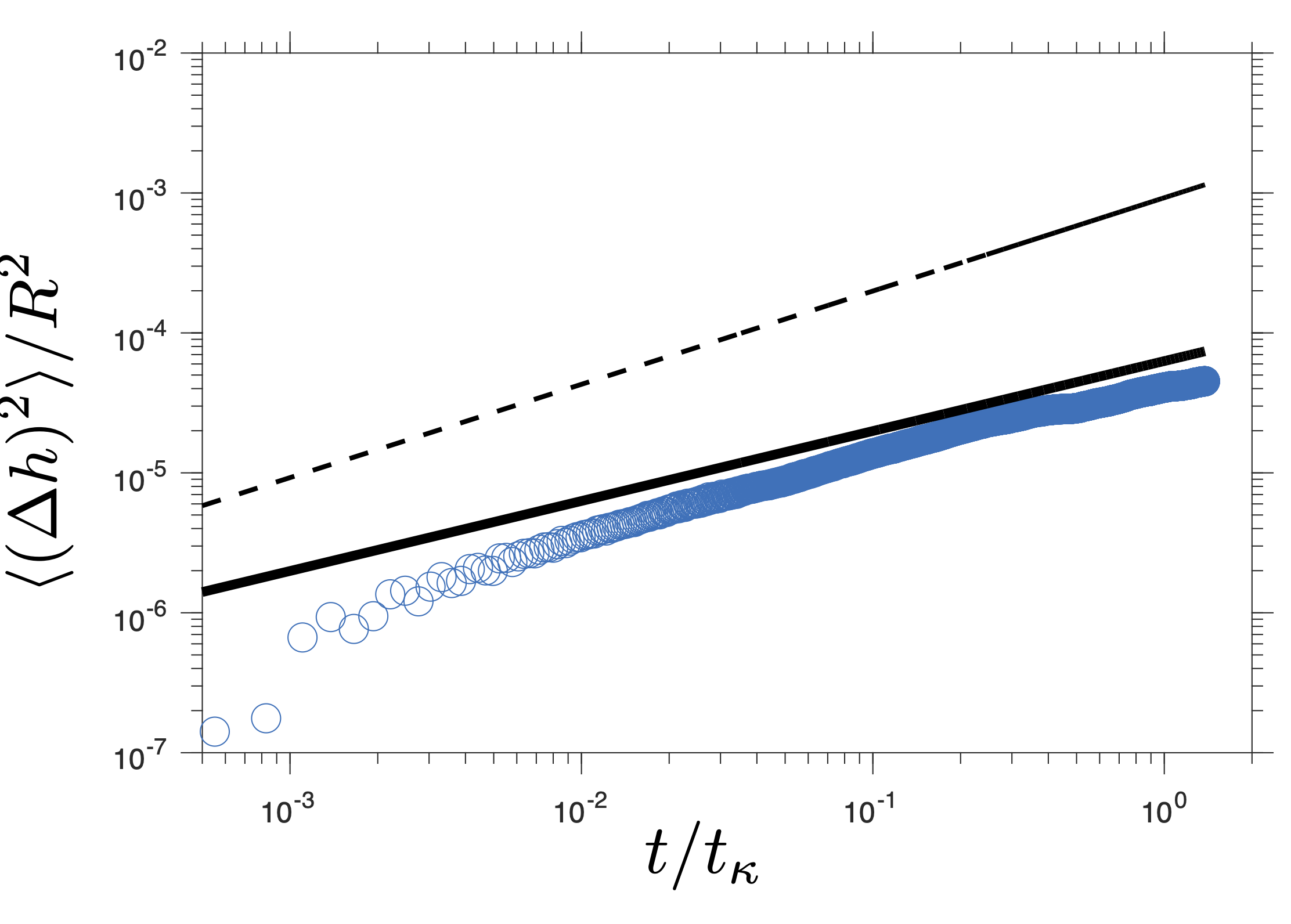}
\centering
\caption{\footnotesize{Single-point membrane mean square displacement (dynamic roughness) of a DPPC:Chol (7:3) membrane. The solid line is the the viscous asymptote \refeq{MSD-asymptotes} with $\chi_s=150$, same as in Figure \ref{fig2}. The dashed line is the non-viscous behavior \refeq{ZG-asymptotes}.}}
\centering
\label{figMSD}
\end{figure}

It follows that the scattering from vesicles in this time range and large scattering wavenumbers, $k R\gg 1$, would still exhibit a stretched exponential DSF, $S(k,t)\approx\exp\left[-\left(\Gamma_k t\right)^{\alpha}\right]$, but with a modified stretching exponent $\alpha$ and relaxation rate $\Gamma_k$, changing  from $\alpha=2/3$ and $\Gamma_k\approx  (\kT)^{3/2}k^3/\kappa^{1/2}\eta$  to $\alpha=1/2$ and
\begin{equation}
\label{VG}
\Gamma_k \approx (\kT)^2k^4 R^2/\kappa \eta_m
\end{equation}
Indeed, the MSD (sometimes termed dynamic roughness) of DPPC:Chol (7:3) bilayer, shown in Figure \ref{figMSD}, follows the viscous behavior predicted by \refeq{MSD-asymptotes}. Notably,  there are no fitting parameters in this plot, as the value for the membrane viscosity is  obtained from the analysis of the Fourier modes ACF shown in Figure \ref{fig2}.

This result may be of relevance to the discussion about the cholesterol stiffening of DOPC lipid bilayers reported from NSE experiments but not found in flickering of giant vesicles \cite{Chakraborty21896, Nagle:2021}. Since the Saffman-Delbrück length even for typical low-viscosity lipid as DOPC is about a micron, the membrane dissipation affects the fluctuations of the submicron liposomes used in the NSE experiments.  Currently, the NSE data is analyzed with the ZG model, which neglects membrane viscosity. Our results suggest that using the ZG model can misinterpret the effect of membrane viscosity as increased bending rigidity. A definitive answer, however, requires generalizing the theory to account for lipid density fluctuations \cite{Seifert-Langer:1993,Watson:2010,Watson:2011,Miao:2002}.

\subsection{Polymer membranes}

The diblock-copolymer membranes display more complex dynamics.
First, the apparent increase in viscosity with mode number is more pronounced and manifests itself in both the short-time and long-time dynamics. In Figure \ref{fig2}cf, the long-time viscosity ranges from $\visrat_s=180$ to 1100 and the short-time viscosity ranges between $\visrat_s=60$ and 280. Second, the viscosity obtained from the short-time asymptote  tends to be lower than the one obtained from the single-exponential long-time ACF fit. Third, the ACF may exhibit multiple exponential decays. The PS1 membranes relax much more slowly compared to the DPPC:Chol (7:3) ones even though the lipid membrane has higher viscosity, because of much smaller bending rigidity (approximately by factor of 6). Accordingly,the long time ACFs may be noisier. However, the mode dependence of the viscosity in the short-time ACFs suggests that additional dissipative mechanisms may be at play.

\begin{figure}[h]
\centering
	\includegraphics[width=\columnwidth]{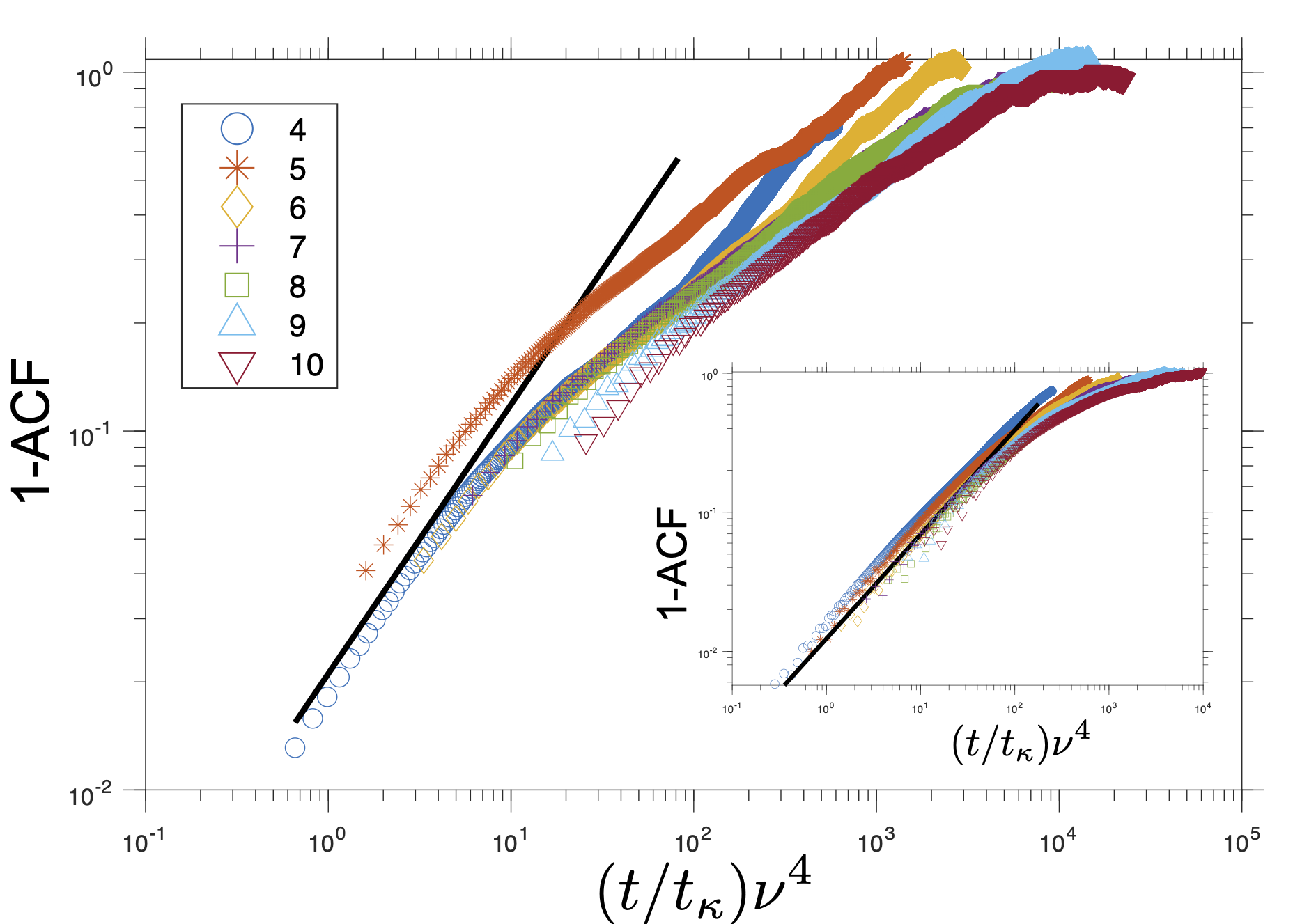}
	\caption{\footnotesize{
 \col{Normalized} mean square displacement for Fourier modes 4-10 of the fluctuating equatorial contour of a polymersome made of PS1 
 plotted as a function of the dimensionless time multiplied by the fourth power of the mode. Inset: Same plot but for DPPC:Chol vesicle showing collapse of the data. The solid line is the viscous asymptote \refeq{ACF-asymp1} for tensionless membranes with $\visrat_s=150$. The time scale $t_\kappa=\eta R^3/\kappa$.
	}}
\label{fig4}
\end{figure}

Figure \ref{fig4} compares the short-time ACFs of the lipid and polymer membranes. Rescaling the time by $\nu^4$, as suggested by Eqs. \ref{ACF-scaling-assumption}-\ref{ACF-scaling-function}, collapses the data for the PC membrane -- especially at short times, as implied by the weak $\nu$ dependence of the scaling function \refeq{ACF-scaling-function} \col{at intermediate and long times} -- confirming that hydrodynamic dissipation in the membrane is solely responsible for the relaxation rate. 
The PS data not only do not collapse but also exhibit crossover to relaxation with a lower exponent. 

The more complex dynamics of the polymer membranes seen in the ACF behavior in Figure \ref{fig4}b may result from several factors. 
In addition to intermonolayer friction \cite{Rumy:2002, Watson:2010,Monroy:2009,Mell:2015, Anthony:2022}, the membrane could be viscoelastic due to the slow polymer chain dynamics. Viscoelastic behavior of entangled and non-entangled polymer systems arises from different stress relaxation mechanisms. However, it is unclear which mechanism is responsible for the relaxation of the bilayer shear stress. The rheology of a sheared polymer brush along the normal ('z'-) axis has been studied extensively (see, e.g., Ref. \cite{Doyle:1998} and refs. therein), yet here we deal with the two-dimensional in-plane shear flow on which much less is known. 
Phenomenologically, if we assume a power-law complex modulus to describe the membrane viscoelasticity, $G_m^{*}(\omega)=G_0(i\omega\tau_m)^{\alpha}$ where $\tau_m$ is a relaxation time, it can be expected (following similar lines as in Ref. \cite{Granek:2011}) that the stretching exponent changes to $\zeta={3\over 4} \alpha$. (In the case of Newtonian viscous fluid,  $\alpha=1$ and we recover $\zeta=3/4$ and the membrane viscosity is $\eta_m=G_0\tau_m$.) While $\alpha=1/2$ is predicted for the high-frequency Rouse chain dynamics, the Rouse time for a chain of $N$ monomers scales as $\sim N^2$, and with $N\sim 40$ it is, therefore, likely orders of magnitude shorter than the experimental time scale, such that the power law $G_m^*(\omega)$ should be absent. Entangled chains take a much longer time to relax the stress they endure. Typically, the chains in the bilayer are expected to be weakly entangled similar to a polymer brush. If they nevertheless do entangle \cite{Hoy:2007}, the long (exponential in $N$) ``arm retraction" time, similar to that appearing in the reptation of star polymers \cite{Doi-Edwards}, might control the stress relaxation time, and this can give rise to a complicated viscoelastic behavior in the observation time scale. Finally, we pose the possibility that the membrane viscosity is wavelength, i.e. mode number $\ell$, dependent, reflecting a more general system-size dependence property. The latter may explain the conflicting membrane viscosity values obtained from small system size simulations \cite{Zgosrki:2019,Fitzgerald:2023} and measurements with giant vesicles \cite{Faizi:2022, Rumy:2002, h-smith2013}. If so, the higher, long-time viscosity, seen in Figure \ref{fig2}bd, reflects the longer wavelength viscosity, and suggests the expected value to be {measured in} electrodeformation and other large-scale flow experiments.

\section{Conclusions and outlook}

We show that membrane viscosity plays a significant role in  the undulation dynamics of quasi-spherical vesicles if the  Saffman-Delbr\"uck length $\eta_m/\eta$ is comparable to the vesicle radius $R$, that is $\visrat_s\gtrsim 1$.
 This can occur if either the membrane viscosity is large, $\eta_m\gtrsim\eta R$, as in diblock-copolymer bilayers or lipid bilayers in the liquid-ordered phase such as DPPC:Cholesterol, or vesicle size is small, $R\lesssim \eta_m/\eta$, as in submicron liposomes.

 The theory predicts that hydrodynamic dissipation in the membrane gives rise to a unique signature in the flickering spectrum of vesicles and liposomes. 
 The time autocorrelation function of the Fourier modes describing the contour fluctuations of the vesicle equatorial cross-section obeys a stretched exponential decay with universal stretching exponent 3/4.
 This new feature combined with the variance of the fluctuations allows to measure  {\it independently} membrane viscosity and bending rigidity from the flickering experiment. This is usually impossible in scattering methods such as neutron spin echo (NSE), where all physical parameters are obtained solely from the relaxation curves.

Applying the new analysis to  flickering experiments of giant vesicles show that  DPPC:Cholesterol membranes behave as a Newtonian fluid. Polymer membranes exhibit more complex rheology, which may arise from viscoelasticity or interleaflet friction. To account for the latter effect, a theory, which would be the analog of the \cite{Seifert-Langer:1993} model for planar bilayer needs to be developed to include membrane viscosity \cite{Miao:2002}. The new theory will also be relevant to NSE experiments \cite{Granek:EPJE}, since $\chi_s$ becomes significant for the submicron liposomes employed by this method and lipid density relaxation due to bilayer slippage is important on the time scales of the curvature fluctuations \cite{Watson:2010,Watson:2011}.
 
Our results highlight that the viscous properties of lipid bilayers significantly affect the bending dynamics of membranes and provide new insights into the dynamical aspect of curvature remodeling. 
The dynamic flickering experiment can serve as a noninvasive tool for the comprehensive analysis of membrane mechanics in vitro. The method can be applied to more complex membranes such as asymmetric bilayers, hybrid membranes made of lipid-polymer mixtures, and charged membranes to emulate the conditions in living and synthetic cells.

\acknowledgements{P.M.V and H.A.F acknowledge financial support by NIGMS award 1R01GM140461.  This research was also supported in part by the National Science Foundation under Grant NSF PHY-1748958. R.G. is grateful to NU's Department of Engineering Sciences and Applied Mathematics for its hospitality during the period when this research was completed.}

\begin{widetext}
\appendix

\section{Flickering spectroscopy: theoretical basis}

 Fluctuation spectroscopy analyzes the thermally-driven membrane undulations of giant unilamellar vesicles.  In essence, a time series of vesicle contours in the focal plane (the equator of the quasi-spherical vesicle) is recorded. The quasi-circular  contour is decomposed in Fourier modes,
\begin{equation}
    r(\phi,t) =  R  \sum_{\nu=-\ell_{\max}}^{\ell_{\max}} u_\nu(t)  e^{\im \nu \phi} =R \Big(1+ \sum_{\jj=0}^{\ell_{\max}} \sum_{m=-\jj}^{\jj} f_{\jj m}(t) {{Y}}_{\jj m} (\pi/2,\phi) \Big).
    \label{Eq-LA10}
\end{equation}
where $R=(3V/4{\pi})^{1/3}$ is the radius of an equivalent sphere with the volume $V$ of the GUV and $\nu$ is the mode number.  In practice, $\ell_{\max}$ is the maximum number of experimentally resolved modes.
The Fourier coefficient for the $\nu$-th mode is then given by 
\begin{equation}
    u_\nu(t) = \frac{1}{2\pi R}\int_0^{2\pi} r(\phi,t)  e^{-i\nu\phi} d\phi = \sum_{\jj =\nu}^{\jj_{\max}} f_{\jj \nu}(t) n_{\jj \nu}P_{\jj \nu} (0) 
    \label{Eq-LA11}
\end{equation}
as all the other terms integrate to zero. In the above equation, we have inserted the definition of the spherical harmonic, 
\begin{equation}
Y_{\jj m}=n_{\jj m} P_{\jj m}(\cos\theta) e^{\im m \phi}\,,\quad n_{\jj m}=\sqrt{\frac{(2\jj+1)(\jj-m)!}{4\pi (\jj+m)!}}
\label{Ylm}
\end{equation}
where $P_{\jj m}(\cos\theta) $ are the associated Legendre polynomials.

The mean squared amplitude of $u_\nu$ is then given by 
\begin{equation}
    \big<|u_\nu|^2\big> =  \sum_{\jj =\nu}^{\jj_{\max}} \sum_{\jj '=\nu}^{\jj_{\max}} \big<f_{\jj '\nu}^* f_{\jj \nu}\big> n_{\jj \nu}n_{\jj '\nu}P_{\jj \nu} (0) P_{\jj '\nu}^* (0).
    \label{Eq-LA12}
\end{equation}
In terms of the spherical harmonic mode amplitudes, the Helfrich Hamiltonian is given by
\begin{equation}
	H=\frac{1}{2}\sum_{\jj,m} (\jj+2)(\jj-1)\Big( \jj(\jj+1) \kappa + \sigma R^2  \Big) |f_{\jj m}|^{2}
\end{equation}
showing that indeed all modes are decoupled from each other. Equipartition theorem then dictates  
\begin{equation}
    \Big<f_{\jj m}^* f_{\jj 'm'}\Big>=    \kT\bigg[ (\jj+2)(\jj-1)\Big( \jj(\jj+1) \kappa + \sigma R^2  \Big) \bigg]^{-1} \delta_{\jj \jj'} \delta_{mm'}
    \label{Eq-LA8a}
\end{equation}
\refeq{Eq-LA12} therefore simplifies to 
\begin{equation}
 \big<|u_\nu|^2\big> =  \sum_{\jj =\nu}^{\jj_{\max}} \big<| f_{\jj \nu}|^2\big>n_{\jj \nu}^2 |P_{\jj \nu} (0)|^2
\label{Eq-LA13}
\end{equation}
or, explicitly,
 \begin{equation}
 \big<|u_\nu|^2\big> =  \kT \sum_{\jj =\nu}^{\jj_{\max}}  \bigg[ (\jj+2)(\jj-1)\Big( \jj(\jj+1) \kappa+ \sigma  R^2 \Big) \bigg]^{-1} n_{\jj \nu}^2|P_{\jj \nu} (0)|^2
\label{Eq-LA14}
\end{equation}

\subsection{Autocorrelation function of the equatorial plane Fourier modes: Asymptotic behavior for tensionless membranes}
The auto-correlation function (ACF) of the equatorial plane Fourier modes is given by
\begin{equation}
	\Big< u_\nu (0) u_\nu^*(t) \Big>= \sum_{\jj =|\nu|}^{\jj_{max}} A_{\ell\nu} e^{-\omega(\jj)\, t}
	\label{SACF}
\end{equation}
where for $\omega(\jj)$ we shall use here  Eq.\ (\ref{eqw}) (see explanation in the main text for its validity for GUVs dynamics), and
\begin{equation}
    A_{\ell\nu} =\Big<  |f_{\jj \nu}|^2 \Big> n_{\jj \nu}^2 \Big|P_{\jj \nu}(0)\Big|^2\;.
\label{Anu}
\end{equation}

Let us consider the short and long time asymptotic behaviors of the ACF. At times $t$ much longer than the longest relaxation time of Eq.\ (\ref{SACF}), $\omega(\nu)^{-1}$, the ACF reduces to a single exponential relaxation
\begin{equation}
	\Big< u_\nu (0) u_\nu^*(t) \Big> \simeq A_{\nu\nu} e^{-\omega(\nu)\, t}
\end{equation}
For short times, $t\ll \omega(\nu)^{-1}$, we consider the $\jj \gg 1$ behavior of the series terms, and approximately evaluate the series by transforming it to an integral. For $\jj\gg \nu$, we have from Eq.\ (\ref{Ylm}) $n_{\jj \nu}^2 \simeq \jj^{1-2\nu} /(2\pi)$, and the associate Legendre polynomials behave as \cite{Gradshteyn-Ryzhik:1980}
\begin{equation}
	P_{\jj \nu}(\cos\theta)\simeq {2\over \sqrt{\pi}} {\Gamma[\jj+\nu+1]\over \Gamma[\jj+3/2] } {\cos \left[ (\jj+1/2)\theta-\pi/4 + \nu\pi/2\right] \over \sqrt{2\sin \theta}} + o(1/\jj)
\end{equation} 
such that for $\theta=\pi/2$ and $\jj\gg \nu$ we obtain
\begin{equation}
	P_{\jj \nu}(0)\simeq \sqrt{{2\over \pi}}\;  \jj^{\nu-1/2} {\cos \left[ (\jj+\nu)\pi/2\right] }\nonumber
\end{equation}
or
\begin{equation}
			P_{\jj \nu}(0)\simeq\sqrt{{2\over \pi}}\; \jj^{\nu-1/2}\times \begin{cases}
		0  & \jj+\nu \text{ is odd} \\
		(-1)^{(\jj+\nu)/2} & \jj+\nu\text{ is even}	
		\end{cases}
\end{equation}
These lead to
\begin{equation}
	n_{\jj \nu}^2P_{\jj \nu}(0)^2\simeq {1\over \pi^2} \times \begin{cases}
			0  & \jj+\nu \text{ is odd} \\
			1 & \jj+\nu \text{ is even}	
		\end{cases}
	\label{Plq-asymptot}
\end{equation}
and the sum in Eq.\ (\ref{SACF}) becomes
\begin{equation}
	\Big< u_\nu (0) u_\nu^*(t) \Big>= {1\over \pi^2}\sum_{\jj =|\nu|,2 }^{\jj_{max}} \Big<  |f_{\jj \nu}|^2 \Big> e^{-\omega(\jj) \,t}
	\label{ACFapprox}
\end{equation}
Transforming the sum to an integral leads to
\begin{equation}
	\Big< u_\nu (0) u_\nu^*(t) \Big>= {1\over 2\pi^2}\int_{\jj =|\nu|}^{\jj_{max}} d\jj\, \Big<  |f_{\jj \nu}|^2 \Big> e^{-\omega(\jj)\, t}
	\label{ACF-int}
\end{equation}
(The prefactor of $1/2$ arises because in Eq.\ (\ref{Plq-asymptot}) the sum is with interval 2.) Finally, we can use in Eq.\ (\ref{ACF-int}) the large $\jj$ limits of $f_{\jj \nu}$ and $\omega(\jj)$, {\it with the tension being neglected},  and assuming $\chi_s\gg 1$
\begin{equation}
	\Big<  |f_{\jj \nu}|^2 \Big> \simeq  {\kT\over \kappa \jj^4}
\end{equation}
and
\begin{equation}
	\omega(\jj) \approx \begin{cases}
		\frac{ \kappa }{4\chi_s\eta R^3}\, \jj^4 & 1 \ll \jj\ll \chi_s\\
		\frac{ \kappa }{4\eta R^3}\, \jj^3 & \chi_s\ll \jj\ll \jj_{max}
	\end{cases}
\label{taul-asymptotes}
\end{equation}
Thus, for $\nu\gg 1$ the variance of the equatorial Fourier modes (i.e., the static, $t=0$, ACF) is evaluated to give
\begin{equation}
	\Big< |u_\nu|^2 \Big>\simeq \frac{1}{6\pi^2}\frac{k_BT}{\kappa\nu^3}
	\label{static-variance-approx}
\end{equation}
which shows the known $\sim \nu^{-3}$ scaling \cite{Pecreaux:2004,Gracia:2010}. Furthermore, in this large $\nu$ limit we numerically find
\begin{equation}
    \frac{A_{\nu\nu}}{\Big< |u_\nu|^2 \Big>}\simeq \text{const.}\,\nu^{-1/2}
\end{equation}
($\text{const.}\simeq 5.32$), determining the ratio of the surviving ACF exponential relaxation amplitude to the initial ACF (i.e., static) value.

We now wish to replace the lower bound of the integral in Eq.\ (\ref{ACF-int}) by 0, however, since the integral diverges as the lower bound approaches $0$, we make use of the identity
\begin{equation}
	\Big< u_\nu (0) u_\nu^*(t) \Big>=  \Big< |u_\nu|^2 \Big>- {1\over 2\pi^2}\int_{\jj =|\nu|}^{\jj_{max}} d\ell\, \Big<  |f_{\jj \nu}|^2 \Big>\left( 1- e^{-t \omega(\jj)}\right)
	\label{autoFqapproxint1}
\end{equation}
Note that the second integral is essentially half of the MSD of the $\nu$-th Fourier mode, $\langle\Delta u_{\nu}(t)^2\rangle\equiv \langle\left(u_{\nu}(t)-u_{\nu}^*(0)\right)^2\rangle$.

``Scaling" the integral in Eq.\ (\ref{autoFqapproxint1}) (i.e. changing the variable of integration), we obtain the two ``short times" regimes, commencing by a solvent viscosity dominated regime which is followed by a membrane viscosity dominated regime. The crossover time separating the two regimes is
\begin{equation}
	t^{*}\approx \frac{4\eta R^3}{\kappa \chi_s^3}\; .
\end{equation}

For the earlier regime of ``short times", $t_0 \ll t \ll t^{*}$, where $t_0$ is the shortest relaxation time, $t_0=1/\omega(\jj_{max})$, we get the solvent dissipation regime
\begin{equation}
	\Big< u_\nu (0) u_\nu^*(t) \Big>\simeq   
	\Big< |u_\nu|^2 \Big>- {1\over 8\pi^2} {k_BT\over \eta R^3}\; t \left[{1\over 3}\ln\left(\frac{4\eta R^3}{\kappa \nu^3\, t}\right) + 0.2838\right] 
	\label{SACF-asymp2}
\end{equation}
For the late regime of ``short times", $t^{*}\ll t\ll \omega(\nu)^{-1}$, we obtain the membrane dissipation regime
\begin{equation}
	\Big< u_\nu (0) u_\nu^*(t) \Big>\simeq   
		\Big< |u_\nu|^2 \Big>- {\Gamma[1/4]\over 6\pi^2} {k_BT\over \kappa} \left(\frac{ \kappa } {4\chi_s\eta R^3}\; t\right)^{3/4} 
	\label{SACF-asymp1}
\end{equation}
presenting a non-exponential relaxation of the ACF from its static value; effectively (to first order) it is a stretched exponential decay with stretching exponent $3/4$. Equivalently, the amplitude of the equatorial Fourier mode $\nu$ is anomalously diffusing with exponent $3/4$, since its MSD is equal to $2\left(\langle |u_\nu|^2 \rangle- \langle u_\nu (0) u_\nu^*(t) \rangle\right)$. This exponent also describes the short-time (polymer segment) anomalous diffusion of semi-flexible polymers obeying the worm-like chain model \cite{Farge:1993,Granek:1997}.

\subsection{Transverse Mean Squared displacement of a membrane segment for tensionless membranes: Implications for the Dynamic Structure Factor}

In Refs. \cite{Zilman-Granek:1996, Zilman-Granek:2002,Granek:1997,Granek:EPJE} it was shown that pure bending undulations, that are dissipated by solvent viscosity, produce a transverse subdiffusion of a membrane segment with a mean squared displacement (MSD) that grows in time as $~t^{2/3}$. At large scattering wavenumbers $k$ that are sensitive to single membrane dynamics, this subdiffusion leads to stretched exponential relaxation, $\sim \text{Exp}\left[-(\Gamma_k t)^{2/3}\right]$, of the dynamic structure factor (DSF) of membrane phases \cite{Zilman-Granek:1996, Watson:2011,Zilman-Granek:2002,Granek:EPJE}. We now discuss how these dynamics are modified when membrane viscosity is included.

\subsubsection{Mean Square Displacement of a membrane segment}

The dimensionless membrane segment MSD at an arbitrary 3D angle $\Omega=(\theta,\phi)$, 	$\langle (\Delta f(t))^{2}\rangle\equiv \langle (f(\Omega,t)-f(\Omega,0))^{2}\rangle$, is given by (note that the MSD with physical dimensions is given by $\langle (\Delta h(t))^{2}\rangle\equiv R^2 \langle (\Delta f(t))^{2}\rangle$ )
\begin{equation}
	\langle (\Delta f(t))^{2}\rangle =
	2k_BT \sum_{\jj =2}^{\jj_{\max}} \sum_{m=-\jj}^{\jj} |Y_{\jj m}(\Omega)|^2 \bigg[ (\jj+2)(\jj-1)\Big( \jj(\jj+1) \kappa+ \sigma  R^2 \Big) \bigg]^{-1} \left( 1- e^{\omega(\jj)\, t}\right)
	\label{segment-MSD}
\end{equation}
Using 
\begin{equation}
\sum_{m=-l}^{l} |Y_{\jj m}(\Omega)|^2 =\frac{2l+1}{4\pi}	
\end{equation}
(which may be verified by using Eq.\ \ref{Ylm}), we arrive at
 \begin{equation}
\langle (\Delta f(t))^{2}\rangle=
{ k_BT\over 2\pi} \sum_{\jj =2}^{\ell_{\max}} (2l+1) \bigg[ (\jj+2)(\jj-1)\Big( \jj(\jj+1) \kappa+ \sigma  R^2 \Big) \bigg]^{-1} \left( 1- e^{-\omega(\jj)\, t}\right)
\label{segment-MSD-1}	
\end{equation}
which is independent of the angle $\Omega$ as expected.

We now assume again {\it vanishing tension}. For times $t_0 \ll t\ll \tau_R$, where $t_0$ and $\tau_R$ are the shortest and longest relaxation times (respectively), $t_0\equiv 1/\omega(\jj_{max})$ and $\tau_R \equiv 1/\omega(2)$, we may use the $\jj\gg 1$ asymptotic of both the spectrum of modes ($\sim \jj^{-4}$) and of the relaxation frequency $\omega(\jj)$, as given by Eq.\ (\ref{taul-asymptotes}), and use $2\jj +1\simeq 2\jj$. We may also replace the sum in Eq.\ (\ref{segment-MSD-1} ) by an integral, with the lower and upper integration limits replaced by $0$ and $\infty$, respectively. This leads to
\begin{equation}
\langle (\Delta f(t))^{2}\rangle\simeq 
{k_BT\over \pi \kappa} \int_{0}^{\infty} d\ell\, \jj^{-3} \left( 1- e^{-\omega(\jj) t}\right)
\label{segment-MSD-2}	
\end{equation}
Note that the latter expression is identical to the one derived for flat membranes using standard 2D Fourier modes \cite{Zilman-Granek:1996, Zilman-Granek:2002}, yet for flat membranes $\chi_s=0$ by definition.

Eq.\ (\ref{segment-MSD-2}) leads to two ``short time" regimes, a solvent viscosity dominated regime that is followed by a membrane viscosity dominated regime. We find

\begin{equation}
		\langle (\Delta h(t))^{2}\rangle\equiv R^2 \langle (\Delta f(t))^{2}\rangle\approx \begin{cases}
		\frac{\Gamma[1/3]}{2\pi 4^{2/3}} \frac{k_B T}{\eta^{2/3}\kappa^{1/3} }\, t^{2/3} & t_0 \ll {t \ll t^{*}}\\
		\frac{1}{4\sqrt{\pi}} \frac{k_BT R}{\sqrt{\kappa \eta_m}}\, t^{1/2} &  t^{*}\ll t\ll \tau_R
	\end{cases}
	\label{SMSD-asymptotes}
\end{equation}
Importantly, $\frac{\tau_R}{t^{*}}\approx \chi_s^4$, which allows for a very wide membrane viscosity dominated regime if $\chi_s\gg 1$.

\subsubsection{Dynamic Structure Factor}


Following Refs. \cite{Zilman-Granek:1996, Watson:2011,Zilman-Granek:2002,Granek:EPJE}, the membrane segment MSD controls the DSF relaxation. Excluding the effect of center-of-mass diffusion, which comes in the form of a multiplying factor, and for polydisperse and (sufficiently) large vesicles -- as described in detail in Ref. \cite{Granek:EPJE} -- the main DSF relaxation is well captured by 
\[
S(k,t)\sim \text{Exp}[-\frac{k^2}{2}\langle (\Delta h(t))^{2}\rangle]\,.
\] 
{{$\langle (\Delta h(t))^{2}\rangle$ is calculated above, for the case of a membrane with relaxed lipid density. 
Our analysis of the fluctuations of a quasi-spherical vesicles shows that}}  in principle the Zilman-Granek (ZG), i.e., bulk viscosity dominated, decay, should be followed by a new, membrane viscosity dominated, decay, as follows
\begin{equation}
	S(k,t)\approx S(k)\times \begin{cases}
		\text{Exp}[-(\Gamma^{(b)}_k t)^{2/3}] & t_0 \ll t \ll t^{*}\\
		\text{Exp}[-(\Gamma^{(m)}_k t)^{1/2}] &  t^{*}\ll t\ll \tau_R
	\end{cases}
	\label{DSF-asymptotes}
\end{equation}
where $\Gamma^{(b)}_k$ is the ZG relaxation rate (possibly with $\tilde\kappa$ replacing $\kappa$ \cite{Watson:2011}), and the new, membrane viscosity controlled, relaxation rate, is given by
\begin{equation}
	\Gamma^{(m)}_k\simeq \frac{(k_BT)^2 R^2}{64\pi \kappa \eta_m} k^4
\end{equation}
Given that $t^{*}$ can be extremely short for viscous membrane vesicles with $R\sim 20-50$ nm, it is quite possible that the whole NSE time range is controlled by membrane viscosity.
Moreover, lipid density fluctuations may influence the dynamics probed by NSE \cite{Watson:2011}. Further efforts are required to include other important effects associated with small vesicle sizes.


 \section{Flickering spectroscopy: experiment} 
 
\subsection*{Vesicle Preparation}
Giant unilamellar vesicles (GUVs) are formed from lipids  such as pure dioleoylphosphatidylcholine (DOPC), stearoyloleoylphosphatdylcholine (SOPC),  mixtures of dipalmitoylphosphatidylcholine (DPPC) and cholesterol (Chol), and polymers such as poly(butadiene)-$b$-poly(ethylene oxide) diblock copolymers, PBd$_{13}$-$b$-PEO$_{11}$ (PS0) and PBd$_{22}$-$b$-PEO$_{14}$ (PS1). The lipids and diblock copolymer were purchased from Avanti Polar Lipids (Alabaster, AL) and Polymer Source Inc. (Montreal, Canada), respectively. The lipid vesicles were produced using the electroformation method \cite{angelova.1987}. The stock solutions are diluted in chloroform to obtain a final concentration of 4 mM. Initially, 7-8 $\mu$l of lipid solution is spread on the conductive side of two Indium tin oxide (ITO, Delta Technologies) glass slides with a 10 $\mu$l gas tight syringes (Hamilton, USA). The slides are placed inside vacuum to evaporate any leftover solvents for at least 3 hours. Afterwards, a 2 mm Teflon spacer is sandwiched between the two glass slides and the chamber is filled with 100 mM sucrose solution. The conductive side of the slides are connected to AC signal generator Agilent 33220A (Agilent, Germany) at a voltage of 1.8 $V_{pp}$ and 10 Hz. The connected chamber is placed inside an oven at $50 ^o$C for 2 hours.  This procedure results in 10-50 $\mu$m sized GUVs. The vesicle suspension is aspirated from the chamber and diluted in 110 mM glucose.  Polymer vesicles were produced with the spontaneous swelling method. Initially, 50 $\mu$l of 6-10 mg/ml (in chloroform) polymer solution was dissolved in 200-300 $\mu$l of chloroform in a 20 ml vial. Polymer films were formed from evaporation by blowing with a nitrogen stream while swirling the solution inside. Afterwards, the vials were dried under vacuum for 2-4 hours. The polymer films were hydrated in the suspending solutions (100 mM sucrose solution) and placed at 60 °C in an oven for 18-24 hours.
\subsection*{Optical microscopy and imaging}
The 
shape fluctuations of a GUV are visualized in phase contrast mode with Axio Observer A1 microscope (Zeiss, Germany). The microscope objectives used are Plan-Apochromat 100x/1.4 Oil Ph3 M27 (FWD=0.17mm), with Immersol 518 F oil,  {{and 63x (0.75 NA) Ph2 (air) }}(Zeiss, Germany). 
Focal depth (FD) or FWHM (full width half maximum) of phase contrast imaging for our setting is determined using the standard formula FD $=\frac{\lambda}{NA^2}$. For a wavelength of transmission light $\lambda=$550 nm, the calculated FD for the 100x observations is 281 nm.

 where $\visrat_s=\eta_m/\eta R$ is the dimensionless membrane viscosity $\eta_m$, $\eta$ is the viscosity the solution inside and outside the vesicle, $E_0$ is the electric field amplitude and $p(\omega)$ is the electric pressure detailed in \textit{Faizi et al.} \cite{Faizi:2022}. The apparent viscosities are measured at different frequencies in the range 0.1-1 kHz. The zero-frequency viscosity is obtained by extrapolating a linear fit of the viscosity vs frequency data.
 Electric field of 8 kV/m produces  a good range of data in the linear initial slope.  
 
\subsection*{Flickering experiment}
 
 Flickering spectroscopy has become a popular technique to extract out membrane rigidity and tension 
due to its non-intrusive nature and well developed statistical analysis criteria. The methodology is detailed in 
 \cite{Gracia:2010,Faizi:2019, Faizi:2020}. 

The 
 bilayer elastic properties, 
bending rigidity and tension, are obtained 
from the variance of the shape fluctuations (\refeq{ACF} at $t=0$).  The resulting spectrum is shown in Fig. \ref{fig1}b. Rescaling the variance by the bending rigidity $\kappa$ collapses the data, see Fig. \ref{fig1}c,  confirming that the static spectrum is dominated by the bending rigidity.


Figure \ref{figSACF}a illustrates the behavior of the ACF of undulations with different wavelengths. The relaxation rate increases with mode number (or equivalently decreases with undulation wavelength), see Figure \ref{figSACF}c, and the corresponding ACF decays faster. The membranes  viscosity decreases the relaxation rate and slows down the ACF decay, see Figure \ref{figSACF}b. Note that the experimentally  obtained ACF may show a drop between  the first two time points due to experimental white noise, see inset in  Figure \ref{figSACF}a, however this does not affect the decay  of the ACFs. Accordingly, all ACFs are normalized by  its value immediately after $t=0$. 

 \begin{figure}[h]
	\includegraphics[width=\linewidth]{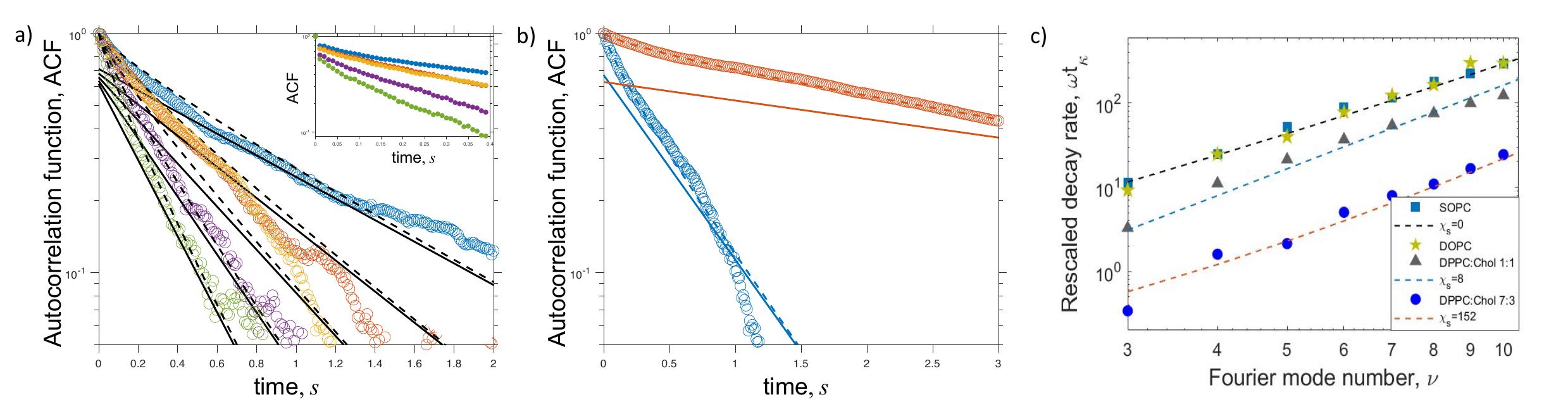}
	\caption{\footnotesize{(a) Autocorrelation functions (ACF) as a function of time (in seconds) for Fourier modes 4-8 of the fluctuating equatorial contour of vesicles made of PS0. Inset: The raw data for the ACF at short times. (b) The ACF of the 6th mode of vesicles made of low viscosity  PS0 ($\chi_s=0$) and high viscosity PS1 ($\chi_s=50$) diblock-copolymer bilayer membranes. Solid lines correspond to the single-exponential long-time behavior, dashed line correspond to the full expression \refeq{ACF}. (c) The long-time single exponential decay rate obtained from the ACF of the phsopholipid DPPC:DOPC membranes as a function of the mode number. Dashed line is the theory \refeq{eqw}. 
	}}
\label{figSACF}
\end{figure}

The integration time effect of the camera is minimized by acquiring images at a low shutter speed of 200 $\mu$s. Images are acquired with SA1.1 high speed camera (Photron) at 50-500 fps for 5-10 mins for a total of 0.1-0.5 million images. Only vesicles with low tension value  in the range $10^{-8}-10^{-10}$ N/m are chosen. This results in a small cross over mode given by  ~$\nu_c=\sqrt{\bar{\sigma}}$ above which the shape fluctuations are dominated by bending rigidity. The ellipsoidal mode ($\nu$ =2) has been ignored from the analysis as it is weighted with most excess area which leads to fluctuations with an increased amplitude.

Correctly resolving the dynamics, especially at long wavelengths, where the effect of the membrane viscosity is expected, requires good statistics. There are two important time scales that affect the quality of the analysis. 
 The first is $t_{\max}$, the  duration of the recording. The second is the time step in the time series, $\dt$, which experimentally corresponds to  the frames per second (fps) or acquisition speed of the movie. $t_{\max}$ is related to  the slowest relaxation mode corresponding to $\jj=2$ (long wavelength curvature fluctuations (smaller mode numbers)  take long time to explore their  configurations), while $\dt$ is determined by  the highest experimentally resolved mode $\jj=\q_{\max}$. We found that a factor of 10 ensures good statistics and converged results. 
 $  t_{\max}=10\omega^{-1}(\ell=2)\,,\quad  \dt=1/10 \omega^{-1}(\ell=\q_{\max})$,
where $\omega$ is given by \refeq{eqw}.

\begin{figure}[h]
\includegraphics[width=\linewidth]{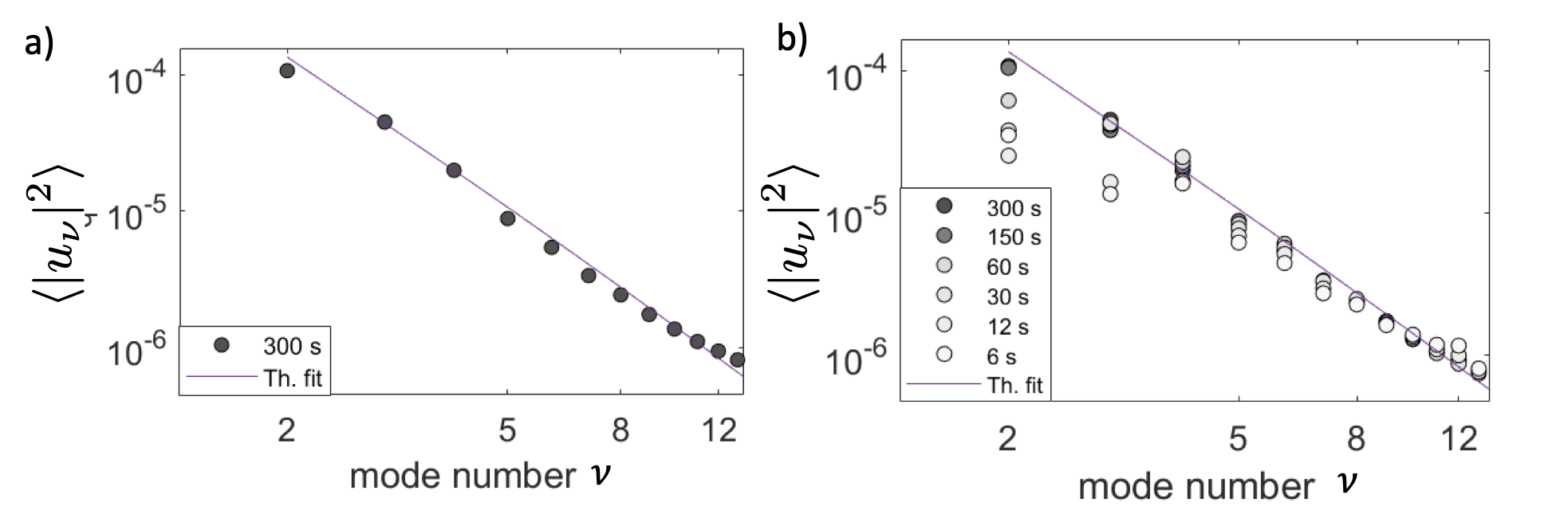}
\centering
\caption{\footnotesize{(a) Equilibrium fluctuation spectrum of SOPC taken over a data set of 300 seconds. b)  The same data set is trimmed to first 150, 60, 30, 12, 6 seconds. For this system $\chi_s\sim 0$ and $t_{\max}=37$s, $\dt=0.0034$s, which corresponds to 294fps. Solid line is the theoretical fit with the static spectrum
}}
\centering
\label{figSopc_eq}
\end{figure}

Figure \ref{figSopc_eq} illustrates the importance of collecting long enough data set 
to ensure good statistics for the equilibrium spectrum. Figure \ref{figSopc_eq}a shows a typical SOPC fluctuation equilibrium spectrum determined from data taken over 300 seconds ($t_{\max}=37$ s). If the same data is trimmed shorter time, e.g., the first 150 seconds only, the same spectrum is recovered. This implies that  experimental convergence has been achieved. However, trimming the data to the first 60 seconds, or shorter, results in artifacts starting from lower mode number spectrum and creeping to higher ones. The spectrum can be misinterpreted with a higher tension value. 
This demonstrates the importance of having data with good and sufficient temporal statistics.

\begin{table}[h]
\caption{\footnotesize{Computed $t_{\max}$ and $\dt$ for membranes with varying dimensionless membrane viscosity, $\visrat_s$. Other parameters assumed are $\kappa=25\kT$, $R=10\,\mu{m}$, $\eta=10^{-3}$ Pa.s, $\sigma=0$. These numbers were also chosen for experiments as well.}}
\centering
\begin{tabular}{|c|c|c|c|c|c|}
\hline
Membrane Viscosity&$t_{relax,2}$ (s)& $t_{\max}$ (s) & $t_{relax,10}$ (s)&$\dt$ (s)&fps\\\hline\hline
$\visrat_s=0$&3.7 &37 & 0.034&0.0034&294\\\hline
$\visrat_s=1$&4.8  &48 & 0.037&0.0037&270 \\\hline
$\visrat_s=10$&14.5 &145 &0.067&0.0067&149 \\\hline
$\visrat_s=100$&112 &1120 &0.36&0.036&27\\\hline
\end{tabular}
\label{timescales}
\end{table}

\begin{figure}[h]
\includegraphics[width=\linewidth]{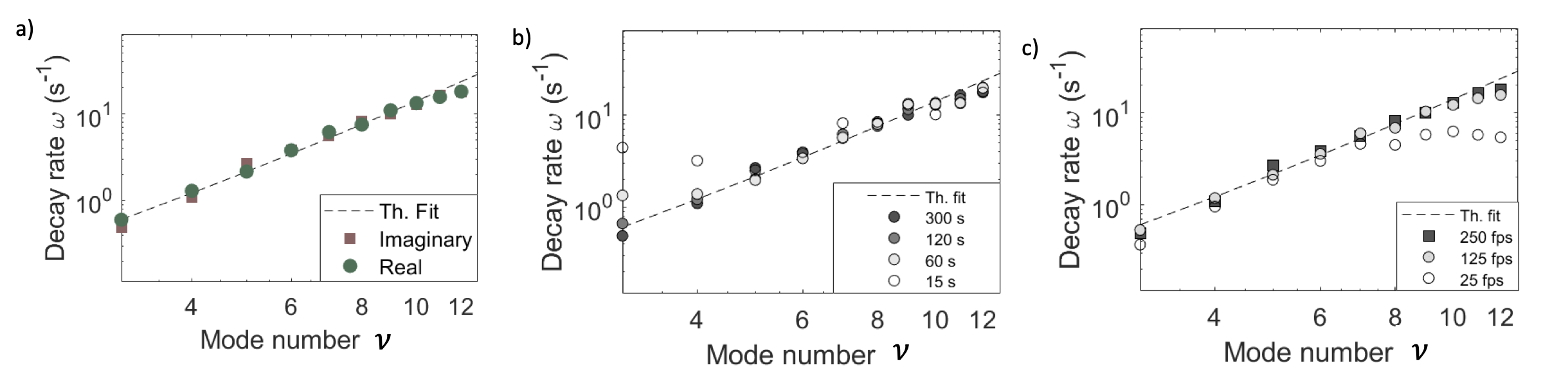}
\caption{\footnotesize{(a) Decay rate obtained from both imaginary and real parts of the Fourier modes for a SOPC vesicle with $\chi_s<1$. The experiment was conducted for 300 secs. b)  The same data set is trimmed to first 120, 60, 15 seconds at 250 fps. c)  The same data set taken for 300 seconds but evaluated at different fps. }}
\label{figSopc_tc}
\end{figure}
\begin{figure}[h]
\includegraphics[width=\linewidth]{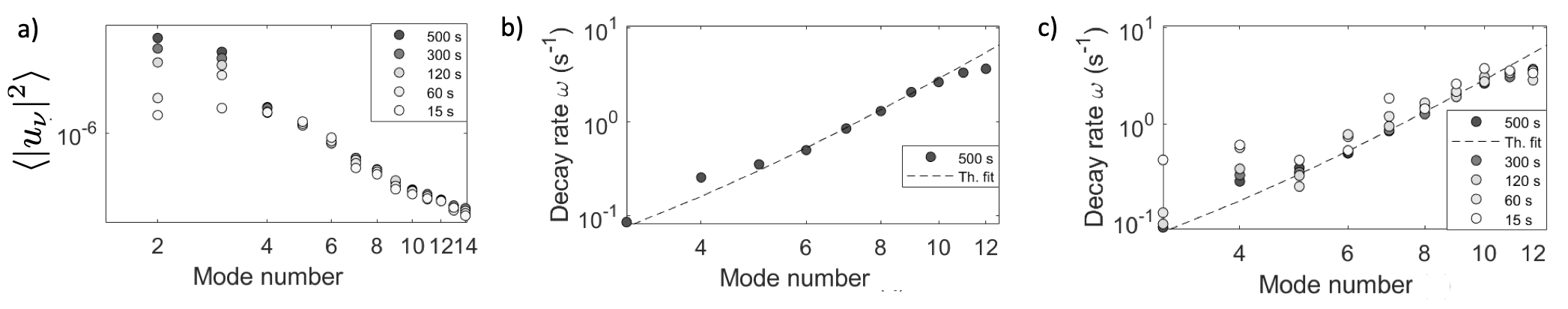}
\centering
\caption{\footnotesize{a) Equilibrium fluctuation spectrum of DPPC:Chol (1:1) taken over a data set of 500 seconds. b) Decay rate obtained for the same vesicle in a) with $\visrat_s\sim10$ c)  The same data set is trimmed to first 300, 120, 60, 15 seconds at 250 fps.}}
\centering
\label{figdppc_tc}
\end{figure}

For time correlations analysis of the same vesicle as in Fig. \ref{figSopc_eq}a, Figure \ref{figSopc_tc} illustrates the importance of time scales $t_{\max}$ and $\dt$. Analyzing the data for a trimmed data set results in artifacts in data interpretation at lower modes as shown in Figure \ref{figSopc_tc}b with a higher membrane tension. Similarly, analyzing the data set at lower fps or lower temporal resolution affects higher mode number data as shown in \ref{figSopc_tc}b. It can been seen that the artifacts exacerbate for membranes with higher membrane viscosity like DPPC:Chol (1:1) with $\visrat_s\sim10$ as shown by time correlation analysis in Figure \ref{figdppc_tc}.

Increasing  the membrane viscosity slows down the dynamics and requires recording vesicle fluctuations over longer times to achieve good statistics. The power-spectrum of DPPC:Chol (1:1) membrane which has  with $\visrat_s\sim10$ is shown in Fig. \ref{figdppc_tc}a.  The time correlations analysis is also very sensitive to the data quality. Fig. \ref{figdppc_tc}b. shows that insufficient data give rise to artificial increase in the decay rates of the lower wave-number modes. Analyzing the data set at lower fps or lower temporal resolution affects the decay rates, of the higher modes,  see  Fig. \ref{figdppc_tc}c.

Table \ref{bending_Table} summarizes the material properties of the bilayer membranes

\begin{table*}[b]
\small
  \caption{\footnotesize{Membrane bending rigidity and viscosity  for various bilayer systems at  25.0 $^o$C.  $L_d$ and $L_o$ refer to liquid-disordered and liquid-ordered phases respectively. FA refers to fluctuation analysis and ED refers to Electrodeformation. $\chi_s$ is computed for GUV with radius 10 $\mu{m}$ in  solution with viscosity $10^{-3}$ Pa.s.m using the membrane viscosity from ED. 
  The data for the bending rigidity and ED viscosity for the PS systems are  from \cite{Faizi:2022}. }}
  \label{bending_Table}
  \begin{center}
  \begin{tabular}{llllll}
      \hline
  Composition& Bending Rigidity $\kappa$ & Viscosity $\eta_m$, ED & Viscosity $\eta_m$, FA&$\chi_s$  \\
  \hline
    &     ($\kT$)& (nPa.s.m) &  (n.Pa.s.m)&$=\eta_m/\eta R$  \\
    \hline
    DOPC&21.7$\pm3.1$&4.1$\pm2.6$&not detectable&0.4\\
   SOPC &25.7$\pm3.6$&9.7$\pm3.0$&not detectable&1.0\\
  DPPC:Chol (1:1) &124.4$\pm14.0$ &57.6$\pm12.6$&89.7$\pm$26.3&5.8\\
 DPPC:Chol (6:4)  &152.6$\pm12.6$&83.6$\pm14.3$&106$\pm$47&8.4 \\
 DPPC:Chol (7:3)  &189.6$\pm17.0$&1450$\pm928$&1777$\pm$682&145 \\
 PBd$_{13}$-$b$-PEO$_{11}$ (PS0) &17.1$\pm1.5$& $14.4\pm4.40$& not detectable&1.4 \\
  PBd$_{22}$-$b$-PEO$_{14}$ (PS1)&31.0$\pm5.1$& $686\pm51.0$ & variable&68.6\\
    \hline
    \centering
  \end{tabular}
  \end{center}
\end{table*}



\end{widetext}

\bibliographystyle{unsrtnat}

\end{document}